\documentclass[aps,prl,reprint,superscriptaddress]{revtex4-1}
\usepackage{amsmath}
\usepackage{subfigure}
\usepackage{epsfig}
\usepackage{pstricks}
\usepackage{overpic}
\usepackage{subfigure}

\begin{document}


\title{Random walks over a super-percolating two dimensional lattice}



\author{Fabrizio Cleri}
\email[]{fabrizio.cleri@univ-lille1.fr}
\affiliation{Universit\'e de Lille I, Institut d'Electronique Micro\'electronique et Nanotechnologie (IEMN), CNRS, UMR 8520, Avenue Poincar\'e BP60069, 59652 Villeneuve d'Ascq, France}


\date{\today}

\begin{abstract}
Two-dimensional networks of ordered quantum dots beyond the percolation threshold are studied, as typical example of conducting nanostructures with quenched random disorder. Theory predicts anomalous diffusion with stretched-exponential relaxation at short distances, and computer simulations on lattices of crossing, straight paths of random length 
confirm such a behavior. Anomalous diffusion is interpreted as resulting from the higher probability of taking straight, or ballistic paths, when the traveled distance is comparable or shorter than the lattice characteristic length. Diffusion turns over to normal for longer traveled distances, whence all paths tend to become equiprobable. 
Such random lattice structures represent a model for realistic quantum dot networks, with potential applications in optoelectronics, photovoltaics or spintronics.
\end{abstract}


\maketitle
\setlength{\parskip}{0 pt}


Network-like structures made from one dimensional (1D) nanowires, nanorods or nanotubes as building blocks, can function both as devices and interconnects, and are thus expected to play a prominent role in next-generation nanotechnology. Recently, the synthesis of several such structures in two or three dimensions (2D, 3D) by different techniques was reported for, e.g., CdS, WO$_3$, InAs, PbSe nanowires.\cite{wang,zhou,dick} Applications can range from electronics,\cite{zhu} to electrochemistry,\cite{wei} to strain monitoring,\cite{ho} and so forth. However, very intriguing 2D networks can also be synthesized by the self-assembly of arrays of quantum dots into superstructures, thus obtaining networks that can range from perfectly ordered to quite disordered, over different length scales.\cite{urban,klapp,vank,kagan} 
Such quantum dot superlattices display peculiar electronic band structures,\cite{delerue} effectively behaving as arrays of pseudo-atoms with discrete states combined into bands. Electron injection, tunnelling and hopping through these nanostructures becomes possible, and such 'metamaterials' are predicted to have a strong potential for optoelectronic, photovoltaic, spintronic applications.

In this Letter, I study the general problem of determining the conductivity of electrons hopping between sites occupied by irregular arrays of conducting 'dots', distributed over a planar region. From a more fundamental viewpoint, such structures are also good candidates for studying anomalous diffusion, because of the mixing of transport pathways with largely different probabilites. Non-Fickian diffusion and stretched-exponential correlation functions often arise as a characteristic feature of transport in strongly inhomogeneous media, in such diverse systems ranging from cell membranes to groundwater flow. \cite{java,toppo,fedotov,baeumer} The underlying physics may be reduced to the random walk over a two-dimensional, multiply-connected lattice containing traps at some sites. In our case, an electron would start from a dot, and keep jumping randomly to neighbouring dots, 
until it attempts jumping to an empty site: this represents a dead end, or a 'trap' for the walking electron. In the following, I will firstly derive an analytic approximation for the probability of traveling a path of given length, based on the diffusion over a set of broken domains. Then, I realize a computer model of the random walk on a super-percolating 2D lattice, by constructing random networks of crossing segments with variable aspect ratio. Computer simulations of the traveled path-length and traveled straight distance allow to deduce the asymptotic behavior of the current in such disordered networks, supporting the theoretical prediction of stretched-exponential anomalous diffusion over distances shorter or comparable to the lattice characteristic length, while diffusion turns over to normal at larger distances.  

Let us start from a simple $N \times N$ square lattice with $z$-fold connectivity. The occupation rate 
must be beyond the percolation threshold to permit long range diffusion. However, the presence of geometrical correlations between the sites (i.e., crossing, straight paths of variable length $Q<N$) makes the percolation threshold to depend on the 'aspect ratio', i.e. the average length of straight paths for a given occupation density. Note that, besides dots hopping, this set up is also representative of the transport across a 2D network of randomly dispersed conducting nanowires or nanotubes, with average sizes smaller than the characteristic length ($\sim \sqrt{A}$ for an occupied  surface area $A$).

The density of electrons is assumed to be small enough to neglect direct collision or trap saturation. Therefore, one may focus on the random walk of a single particle. As usual in this kind of problems,\cite{ryazanov, balagurov} we look for the probability $\overline{W}(\textbf{r},t)$ that after $t$ steps the walker is at the lattice position $\textbf{r}$. The equation of motion for the probability is:
\begin{equation}
\overline{W}(\textbf{r},t+1) = \sum_{\textbf{r'}}  p(\textbf{r} - \textbf{r'}) (1 - \delta_{\textbf{r'}})\overline{W}(\textbf{r'},t) + \delta_{\textbf{r}} \overline{W}(\textbf{r},t)
\end{equation}

\noindent
$p(\textbf{r}-\textbf{r'})$ is the (geometric) probability of transition from a site $\textbf{r'}$ to a neighbouring site $\textbf{r}$ in the dense lattice (without traps, or dead-ends); for a fixed lattice geometry and connectivity it is just a constant, equal to some effective 'diffusion coefficient' $D$, the $p$ are inversely proportional to the lattice coordination $z$, and the sum of all the $p$ must equal unity. The switch function $\delta_{\textbf{r}}$ is 0, except at those sites $\textbf{r}$ representing a trap, where it is equal to 1 (therefore arresting the random walk); in practice, it is the matrix of lattice sites occupied by a dot (=0) or empty (=1). The last term in the equation is the (final) contribution of a trapped particle to the equation of motion. From this writing, it is seen that the probability $\overline{W}(\textbf{r},t)$ is indeed independent of the time (conservation of the norm). By assuming a finite mobility, the time variable can be made to correspond with a traveled path length $Q=vt$ at constant velocity $v$.

In fact, we are interested only in knowing the values of the probability $\overline{W}(\textbf{r},t)$ for the sites with $\delta_{\textbf{r}}= 0$, i.e. for the probability of a walker $\textit{not}$ falling in a trap. Then, by summing up over all the paths not ending in a trap, an estimate can be obtained of the probability for an electron to travel over a distance, as a function of the lattice connectivity. Therefore, it is useful to introduce the auxiliary probability:
\begin{equation}
W(\textbf{r},t) = (1 - \delta_{\textbf{r}})\overline{W}(\textbf{r},t)
\end{equation}

This coincides with $\overline{W}$ at the filled sites, and vanishes at trap (empty) sites. By following \cite{ryazanov, balagurov}, its equation of motion is obtained by multiplying the (1) by $1 - \delta_{\textbf{r}}$:
\begin{equation}
W(\textbf{r},t+1) = D \sum_{\rho} \eta( \textbf{r}, \textbf{r}+\rho ) W(\textbf{r}+\rho,t)
\end{equation}

\noindent
where $\eta(\textbf{r},\textbf{r}+\rho)=1$, if both $\textbf{r}$ and $\rho$ are occupied, and 0 otherwise. It is a standard procedure to obtain the formal solution to the (3) above by setting it as an eigenvalue problem:
\begin{equation}
D \sum_{\textbf{r'}} \eta( \textbf{r}, \textbf{r'}) W(\textbf{r}+\rho,t) \phi_n(\textbf{r'}) = \lambda_n \phi_n(\textbf{r})
\end{equation}

As shown in \cite{balagurov}, the general solution is expressed in terms of the initial distribution $W_0(\textbf{r})$ as:

\begin{eqnarray}
W(\textbf{r},t) =\sum_n A_n \phi_n(\textbf{r}) \lambda_n^t \\
A_n = \sum_{\textbf{r}} \phi_n^*(\textbf{r}) W_0(\textbf{r})
\end{eqnarray}

Note that at $t=0$ the (5) becomes an identity because of the orthonormality of the $\phi_n$. Balagurov and Vaks presented a general solution for this problem by 
a spectral method, firstly introduced by Lifshitz \cite{lifshitz}.

However, an easier way to understand the behaviour of the solutions is to firstly note that eq. (3) in \textit{one} dimension reduces to the ordinary diffusion equation over a piecewise connected domain of total length $L$. The traps are represented by $i$ randomly distributed empty sites, corresponding to a concentration $c$. Each trap $i$ delimits a portion on the segment of length $l_i = |x_i-x_{i-1}|$, with $x_0$ the origin and $x_{c+1}=L$. The lengths $l_i$ can be arbitrarily distributed, to reflect the presence (or absence) of spatial correlations in the lattice (1D domain in this case). The probability is subject to $W(x_i,t)=0$ for $i=0,...c+1$ at any $t$, and $W(0<x<L, t=0)=1/L$. In this case, the general solution would be the well known \cite{crank}:
\begin{equation}
W(x_i,t) = (4/L) \sum_n \exp(-k_n^2 Dt/2) \frac{\sin k_n(x-x_i)}{k_n l_i}
\end{equation}

\noindent 
for $k_n=(2n+1)\pi/l_i$.

The probability of traveling freely over a time $t$ (or a total path length $Q=vt$) is the average of integrals of $W(x,t)$ over the segment lengths delimited by the random distribution of traps:
\begin{equation}
P(t) = \sum_i < \int_{x_i}^{x_{i+1}} W(x,t) dx >
\end{equation}

\noindent
the $<...>$ indicating averaging of all possible distribution of the segment lengths $l_i$. 

The term $k_0$ in the sum (7) defines the smallest wavenumber of the walker, whose inverse length defines the largest size, surface area, or volume (in 1D, 2D or 3D) within which the walker will \emph{not} be captured by a trap. One can exploit the analogy between the 1D solution of the diffusion equation on a piecewise continuous domain (7), and its 2D (or 3D) analogous, by replacing the limiting segment by a limiting circle (or sphere). 


By restricting to $k_0$=$\pi/l$, and taking a Poisson distribution of traps, $p(S)$=$c \exp(-cS)$, with concentration $c$, over a circle of surface $S=4/k_0^2=4\pi l^2$,  the average of $W$ is given by:
\begin{equation}
P(t) = \frac { \int_{0}^{\infty} p(S) W(S,t) dS }{ \int_{0}^{\infty} p(S) dS } \propto \int_{0}^{\infty} c e^{-cS} e^{-\pi^2Dt/S} dS
\end{equation}

\noindent
that is, an integral of the type:
\begin{equation}
I(t) = \int_{0}^{\infty} c e^{- cS - \pi^2Dt/S} dS = 2\pi (Dct)^{1/2}K_1(2\pi \sqrt{Dct})
\end{equation}

\noindent
with $K_1$ the modified Bessel function. At long times, $K_1 \sim \exp(-2\pi \sqrt{Dct})/(Dct)^{1/4}$, therefore $P(t)$ decreases as a stretched exponential :
\begin{equation}
 P(t) \sim (Dct)^{1/4} e^{-2\pi(Dct)^{1/2}}
 \end{equation}

On the other hand, for a flat distribution $p(S)=1/c$ (a 'gas-like' distribution of traps) the long-times solution goes rather as a standard (diffusion-like) exponential:

\begin{equation}
 P^*(t) \sim e^{-(\pi c)^2Dt}
 \end{equation}

The above results for the probability of free-travel time qualitatively coincide, apart from numerical factors of order 1, with the results of Balagurov-Vaks \cite{balagurov} and Ryazanov \cite{ryazanov}. 
Notably, sublinear diffusion near the percolation threshold has been often invoked to characterize the random walk over a complex (heterogeneous) configuration space.\cite{gefen,botet,lemke}

\begin{figure}[t]
\begin{center}
\includegraphics[width=2.7cm]{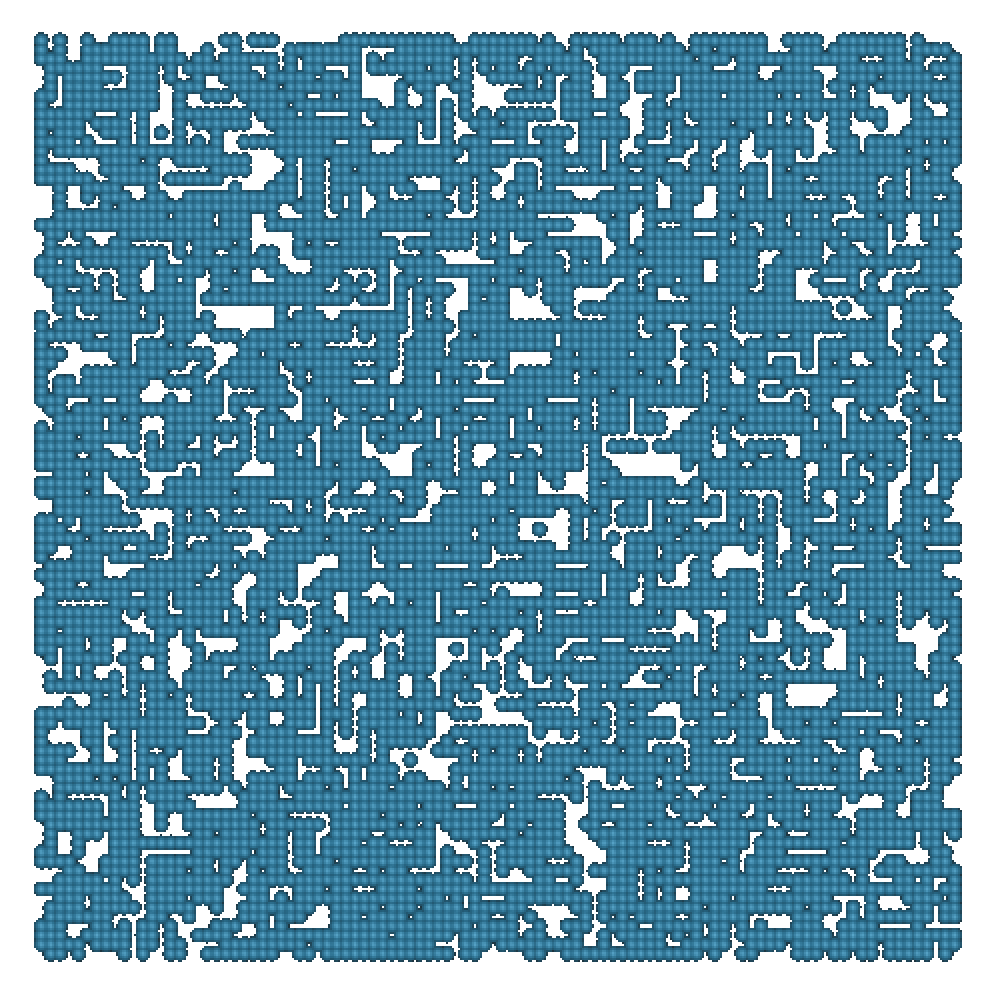}
\includegraphics[width=2.7cm]{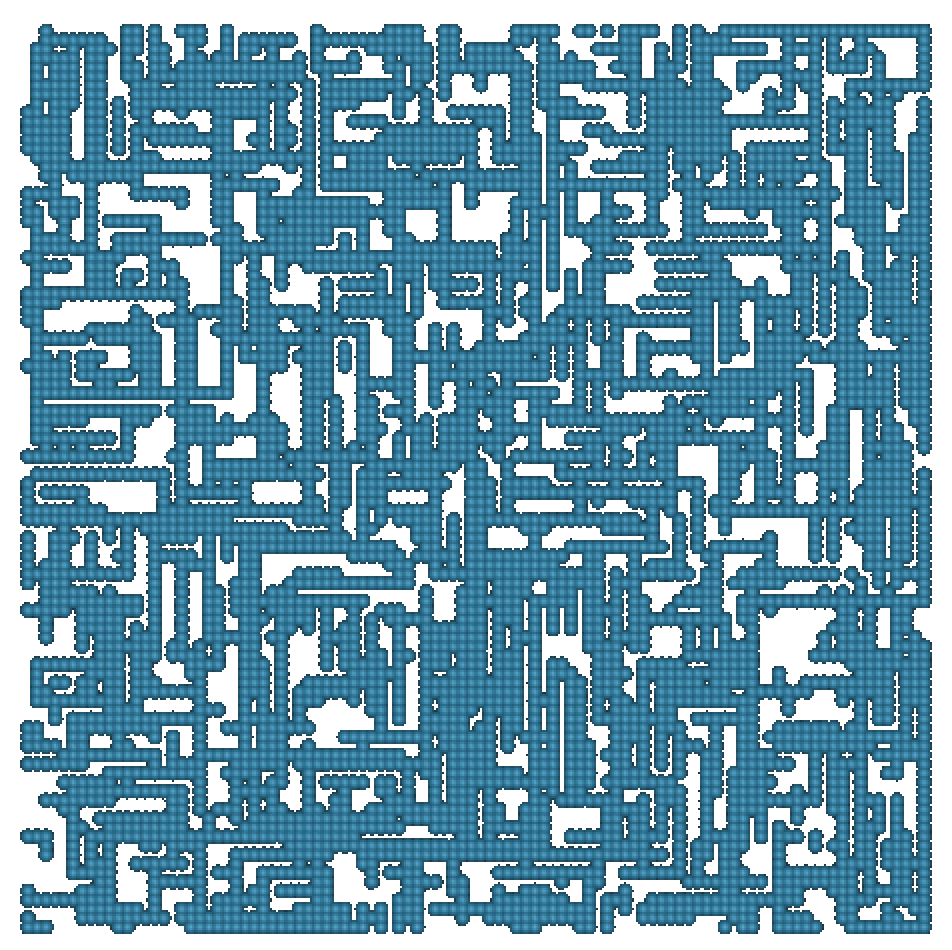}
\includegraphics[width=2.7cm]{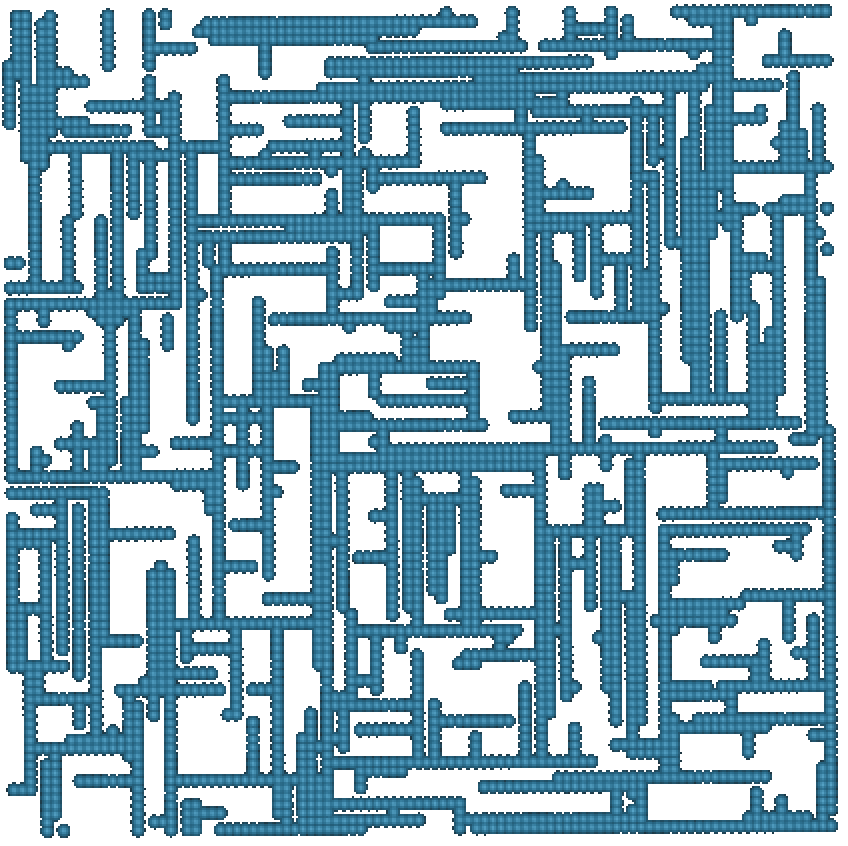}
\includegraphics[width=8.5cm]{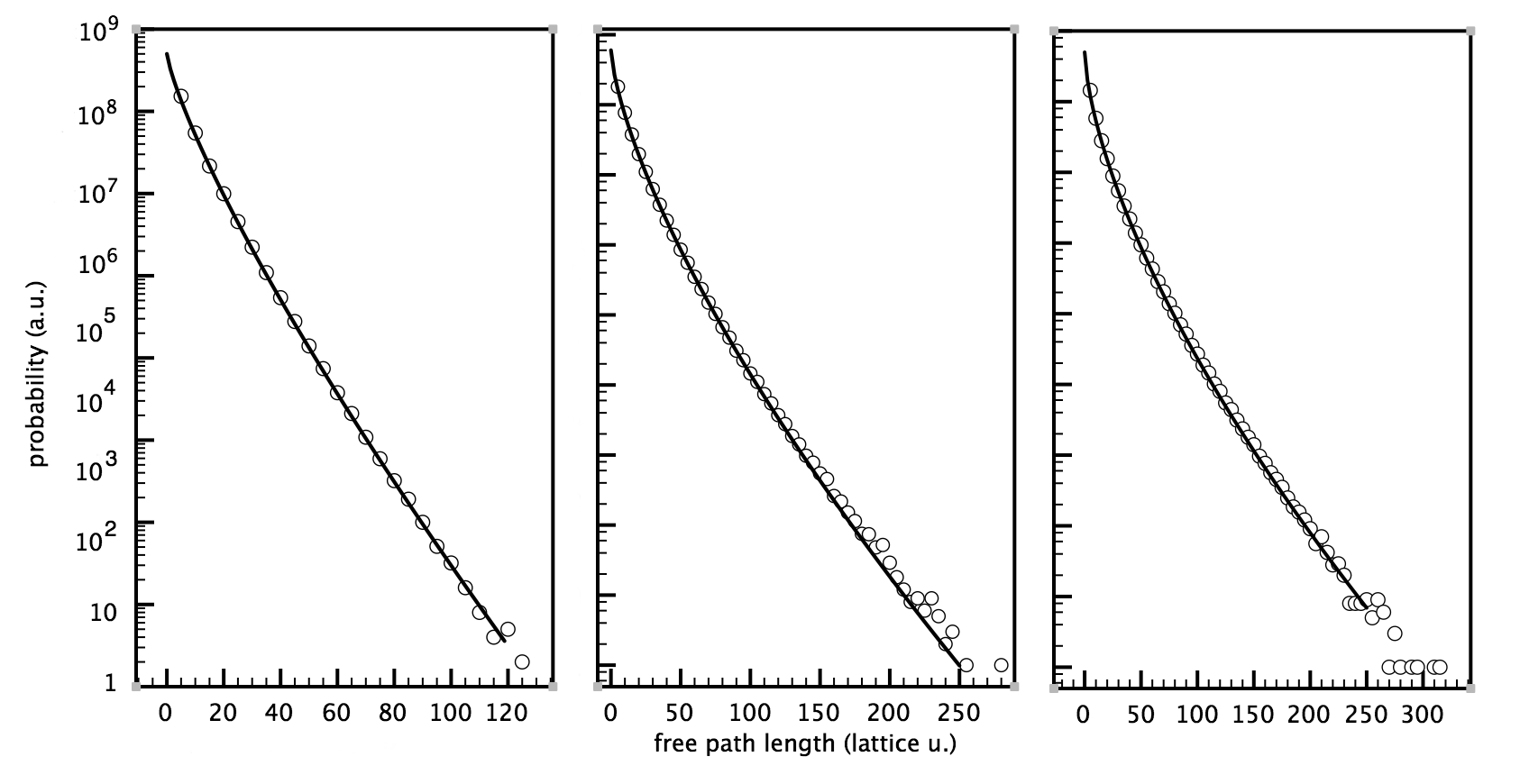}
\caption{Top row: Schematic representation of a portion of two-dimensional $500\times 500$ square lattices, filled with straight segments of aspect ratio $\bar{a}$=6 (left, in lattice units), 12 (middle), and 18 (right). Bottom row: Plot of the path length probability $P(Q)$ for the three values of the segment aspect ratio $\bar{a}$=6 (left), 12 (center), 18 (right). Continuous lines represent fits with stretched-exponential law, with exponents $\alpha$=0.8, 0.7, 0.65, respectively.}
\end{center}
\end{figure}

To verify the above asymptotic limits, I set up a simulation model by filling up a square $N \times N$ lattice with straight segments made up of rows of 'dots'. The lattice has a typical size of $N$=500. 
Using an exponential filling probability, straight segments of dots 
along the $x$ and $y$ direction can be built, with variable segment length $a$. 
For a short average segment length $\bar{a}$ (or 'aspect ratio') the segment-length distribution is flat, while it becomes increasingly closer to a Poissonian as $\bar{a}$ approaches the lattice length size. Therefore, by continuously varying the aspect ratio of the segments, the range from a flat, to a fully Poisson distribution can be explored. Three examples of lattices built with this procedure are shown in Figure 1 (top row): such configurations could be taken as an idealization of real experimental structures, see e.g.  Fig. 1 from ref.\cite{vank}. 

A random walker starts from an occupied site, and proceeds by jumping to neighbour occupied sites, until a move brings it to an empty site (a 'trap'), at which point the walk stops, and a new walker is launched. Typical simulation runs are realised with $10^6 - 10^7$ walkers. Periodic boundaries are applied throughout. In this way, a statistics about the travel free time, and therefore the traveled free path length $Q$ can be accumulated, for each given occupation density and segment aspect ratio. In Figure 1 (bottom row), I show the results for $P(Q)$ for different aspect ratios $\bar{a}$=6, 12, 18 (in units of the lattice mesh), together with stretched-exponential fits of the type $P(Q)=A\exp(-BQ^\alpha)$. The best fit for the stretching exponent gives $\alpha$=0.8, $\alpha$=0.7, $\alpha$=0.65, for the three aspect ratios respectively. Hence, it appears that geometrical correlations in the segment length allow to numerically span the range of probability distributions analytically determined by the extremes (11) and (12), approaching an exponent of 1 as in Eq.(12) for the shorter aspect ratio (in the limit of $a \rightarrow 1$, the gas-like random distribution is recovered), and an exponent of 0.5 as in Eq.(11) for an increasing aspect ratio.


\begin{figure}[t]
\includegraphics[width=8cm]{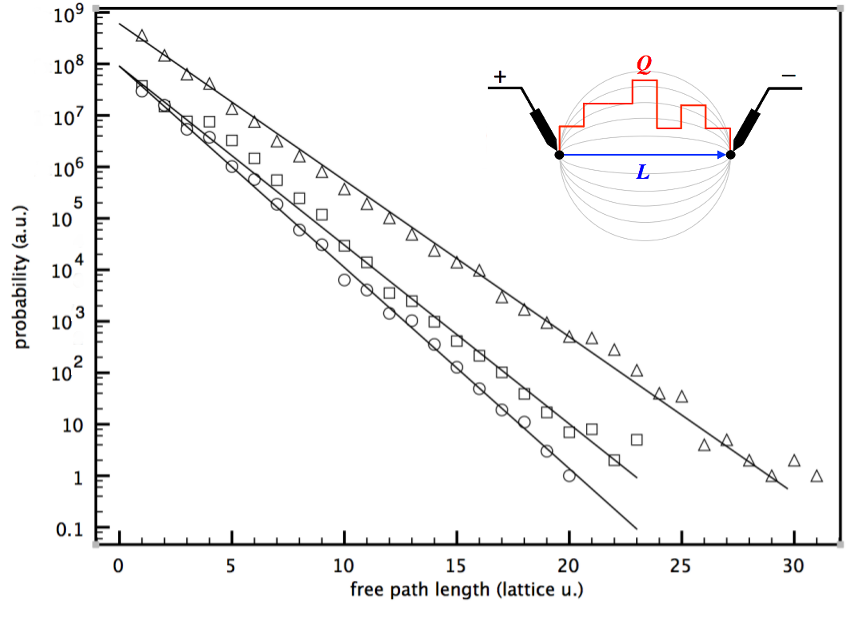}
\caption{Plot of the traveled distance probability $P(L)$ for the three values of the segment aspect ratio $\bar{a}$=6 (circles), 12 (squares), 18 (triangles). Continuous lines represent exponential law fits, with coefficients $\beta$=0.9, 0.8, 0.7, respectively. The inset shows the difference between the path length $Q$ (red) and the traveled distance $L$ (blue), which may be measured under the application of a potential at two points in the lattice. The shape of the electric field lines is depicted in grey.}
\end{figure}

If now, rather than on the total traveled path length, we focus on the traveled \emph{distance}, that is the straight distance $L$ between the end-points of each free path, the plots show in Fig. 2 are obtained. This probability distribution is calculated by adding all the contributions from any path leading to a same value of $L$, for a random distribution of starting points. The distribution $P(L)$ is clearly exponential, $P(L) \sim \exp(-\beta L)$ with a coefficient proportional to the effective diffusion coefficient, $\beta$=0.9, 0.8, 0.7, respectively, for the aspect ratios $\bar{a}$=6, 12, 18. 

Note that the traveled distance is the important quantity when looking at the particle current. For example, in the case of electrons jumping through charged dots, the driving force to push the electrons from one point in the lattice to another one at a distance $L$ would be provided by an electric field. Experiments of such kind may be performed by placing nano electrodes at two contact positions separated by $L$ (see inset in Fig. 2), and shooting a voltage difference between the two tips.\cite{grand} Under such a condition, electrons will flow by taking all the possible paths of length $Q$, leading from 0 to $L$, and the current will result from a weighted average over all such paths.

\begin{figure}[t]
\includegraphics[width=9cm]{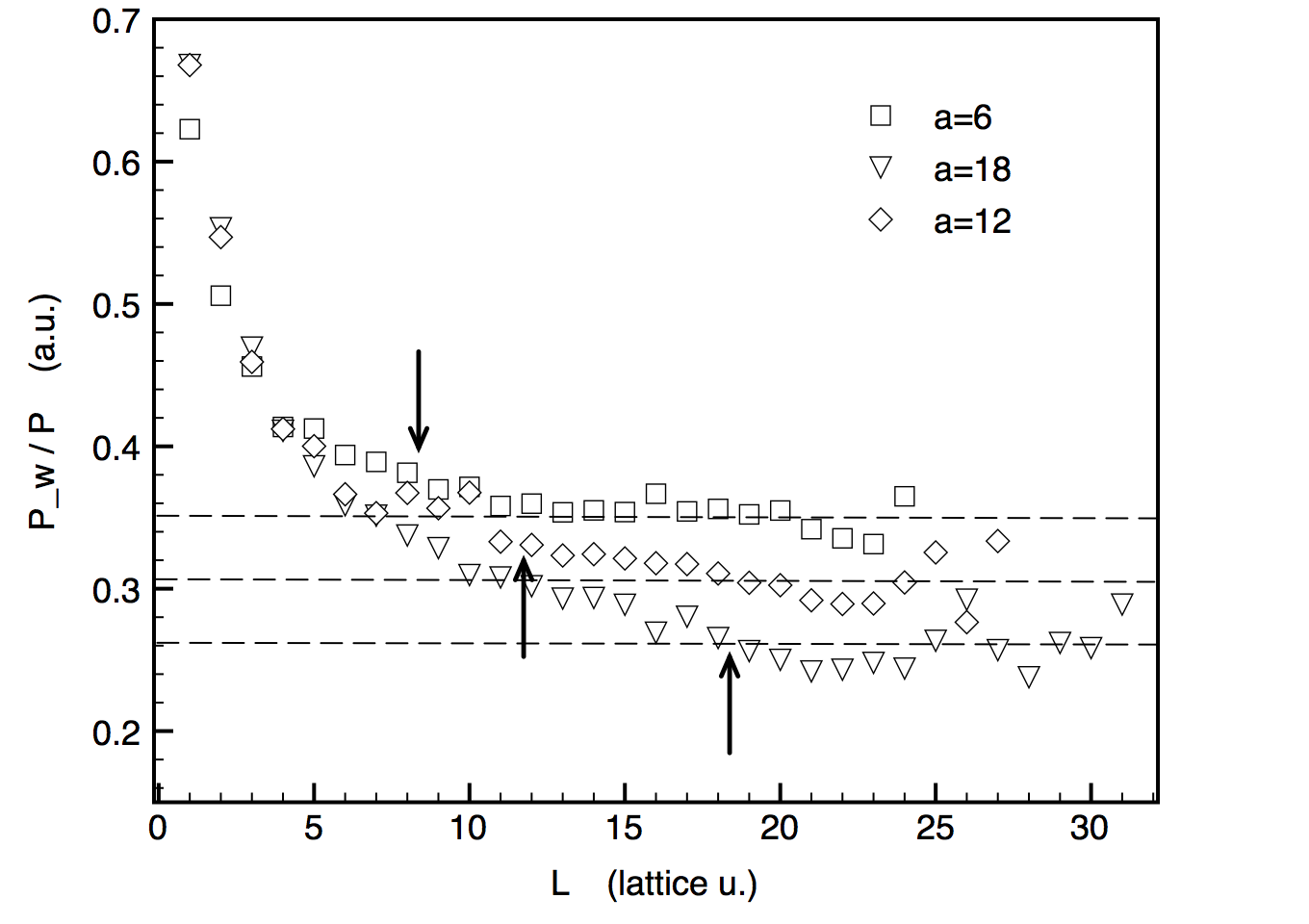}
\caption{Plot of the ratio of field-weighted probability to traveled distance probability $P_w(L)/P(L)$ for the values of the segment aspect ratio $\bar{a}$=6 (squares), 12 (diamonds), and 18 (triangles). Dashed lines represent the respective asymptotic constant values at large $L$. Arrows indicate the approximate value of $L_{sat}$ at which saturation to a constant value occurs.}
\end{figure}

A probability function for electrons traveling the different paths leading to the same distance under a driving force, can be constructed by observing that the electric field between two points at distance $L$, and at different potential, roughly decreases as $L/Q$ (see again inset to Fig. 2). 
This probability $P_w(L)$ can be calculated by weighting each contribution to $P(L)$ by the factor $w=L/Q$. To normalise the result to one electron, I plot in Figure 3 the ratio $P_w(L)/P(L)$ for the aspect ratios $\bar{a}$=6, 12, 18. It can be seen that at values $L$ large compared to $a$, the weighted probability per electron saturates to a constant value, whereas at shorter distances (i.e., a closer distance between the tips of the nano electrodes) the current probability increases faster than linearly. 
At short distances, most paths are straight or nearly straight, i.e. ballistic rather than diffusive. The fact that at large distances, $L>>\bar{a}$, the current (flow probability per electron) becomes constant might be understood by thinking that, once $L$ is larger than the average segment length, no straight paths from 0 to $L$ are possible, and at increasing $L$ all the long electron paths with $Q>>L$ (diffusive) tend to become equiprobable. It can be also observed (see the arrows in Fig. 3) that the value of $L_{sat}$ at which the saturation to a constant value occurs, corresponds quite well to the aspect ratio of the segments filling the lattice (i.e., the average length of the straight segments), $L_{sat} \sim \bar{a}$; the asymptotic saturation value, instead, scales roughly linearly with the aspect ratio.

The above results lead to the following conclusions. First, the stretched exponential behavior is not merely a convenient fitting function, but arises naturally from the distribution of free-path segment lengths. Whenever straight, or nearly straight, paths connecting two points in a disordered network of conductors are available to carriers, these will travel the distance in a time shorter than the average classic (Fickian) diffusion time. This leads to a slower decay of the distance autocorrelation function, a phenomenon often observed near the percolation threshold for various physical systems. Second, the current measured between two random sites in a disordered network of conductors can increase faster than linearly, if the average length of the conducting elements (such as nanowires, nanotubes, arrays of conducting dots) is comparable or larger than the distance between the two points. In other words, the conductance between the two points becomes a nonlinear function of the distance, because of the relative dominance of ballistic over diffusive pathways at short distances.


\begin{thebibliography}{99}

\bibitem{wang} D. H. Wang et al., Chem. Mater. \textbf{18}, 4231 (2006)
\bibitem{zhou} J. Zhou et al., Adv. Mater. \textbf{17}, 2107 (2005)
\bibitem{dick} K. A. Dick et al., Nano Lett. \textbf{6}, 2842 (2006)
\bibitem{zhu} J. Zhu, Nano Lett. textbf{7}, 1095 (2007)
\bibitem{wei} C. Wei et al., Sci. Rep. \textbf{3}, 2193 (2013) 
\bibitem{ho} X. Ho, C. K. Cheng, J. N. Tey, J. Wei, Nanotechnology \textbf{29}, 195504 (2015)
\bibitem{urban} J. J. Urban et al., Nature Mat. \textbf{6}, 115 (2007) 
\bibitem{klapp} F. Klappenberger et al. Phys. Rev. Lett. \textbf{106}, 026802 (2011) 
\bibitem{vank} W. H. Evers, et al. Nano Lett. \textbf{13}, 2317 (2013) 
\bibitem{kagan} S. J. Oh et al., Nano Lett. \textbf{14}, 1559 (2014) 
\bibitem{delerue} E. Kalesaki et al., Phys. Rev. B \textbf{88}, 115431 (2013) 
\bibitem{java} M. Weiss, M. Elsner, F. Kartberg, T. Nillson, Biophys. J. \textbf{87}, 3518 (2004)
\bibitem{toppo} F. Hoefling, T.Franosch, Rep. Prog. Phys. \textbf{76}, 046602 (2013)
\bibitem{fedotov} S. Fedotov, G. N. Milstein, M. V. Tretyakov, J. Phys. A: Math. Theor. \textbf{40}, 5769 (2007)
\bibitem{baeumer} B. Baeumer, Y. Zhang, R. Schumer, Ground Water \textbf{53}, 699 (2015)
\bibitem{ryazanov} G. V. Ryazanov, Teor. Mat. Fiz. \textbf{10}, 271 (1972) 
\bibitem{balagurov} B. Ya. Balagurov, V. G. Vaks, Zh. Eksp. Teor. Fiz. \textbf{65}, 1939 (1973) 
\bibitem{lifshitz} I. M. Lifshitz, Usp. Fiz. Nauk \textbf{83}, 617 (1964) 
\bibitem{crank} J. Crank, \emph{The Mathematics of Diffusion}, 2nd ed., Clarendon Press, Oxford (1975)
\bibitem{gefen} Y. Gefen, A. Aharony, and S. Alexander, Phys. Rev. Lett. \textbf{50}, 77 (1983)
\bibitem{botet} I. A. Campbell, J. M. Flesselles, R. Jullien, and R. Botet, J. Phys. C \textbf{20}, L47 (1987)
\bibitem{lemke} N. Lemke, I.A. Campbell, Phys.Rev. E \textbf{84}, 041126 (2011)
\bibitem{grand} B. Grandidier, private communication

\end{thebibliography}
\end{document}